\documentclass[
    ,final            % use final for the camera ready runs
  ]
  {aipproc}

\layoutstyle{6x9}

\newcommand{\eos}{equation of state}

\newcommand{\be}[1]{\begin{equation}\label{#1}}
\newcommand{\ee}{\end{equation}}
\newcommand{\eg}  {\it e.g.}        % exempli gratia {f.eks.}
\newcommand{\ie}  {\it i.e.}        % id est {d.v.s.}
\newcommand{\etc} {\it etc.}

\begin{document}

\title{Improved phenomenological equation of state
 in the chemical picture}

\author{Regner Trampedach}{
  address={Mt. Stromlo Observatory, Australian National University}
}

\begin{abstract}
I present an overview of an equation of state, being developed in the
chemical picture, and based on the very successful MHD equation of state.
The flexibility of the chemical picture combined with the free-energy
minimization procedure, makes it rather straight-forward, albeit laborious,
to include new effects in the model free-energy, simply by adding new terms.

The most notable additions to the original MHD equation of state, are
relativistic effects, quantum effects, improved higher order Coulomb terms
and a long list of molecules other than the H$_2$ and H$_2^+$ treated so far.
\end{abstract}

\maketitle

%%%%%%%%%%%%%%%%%%%%%%%%%%%%%%%%%%%%%%%%%%%%
%% MAINMATTER
%%%%%%%%%%%%%%%%%%%%%%%%%%%%%%%%%%%%%%%%%%%%

\section{Introduction}

Our understanding of stellar structure and evolution is the foundation for
most astrophysical endeavors, whether related to clusters of stars, galaxies,
the early universe or the distance scale. The quality of stellar modeling
therefore has an impact beyond its own field. Our stellar models, in turn,
depend crucially on the physics they are based on, of which there are two
main categories; dynamical processes, {\eg}, convection and rotation, and
micro-physics, {\eg}, the \eos, opacities and nuclear reaction rates.

The late 80'es and early 90'es saw great advances in {\eos} and opacity
calculations for astrophysical plasmas and the published tables
\cite{mhd4,seaton:opac,rogers:OPAL-EOS,fr:OPAL-opac}
were immediately put to use. The improved atomic physics resolved some
longstanding problems, {\eg}, the disagreement between the theoretical
evolutionary and observed dynamical masses of Cepheids and RR Lyraes
\cite{andreasen:z-opac,double-mod:1988,Yi:OPAL-effect-on-RR-Lyr}.

At the same time, the increase in computer-power also enabled the emergence
of realistic, 3-dimensional simulations of convection \cite{aake:comp-phys},
convection being the most important of the dynamical processes.
Due to the time-scales involved these simulations are constrained to the
outer layers of stars. In the deeper layers, however, the stratification of
the convection zone is adequately described by the mixing-length formulation
\cite{boehm:mlt}. The main parameter, $\alpha$, determining the asymptotic
adiabat of the convection zone, is not prescribed by this formulation. The
3D simulations can be used to calibrate $\alpha$, and to describe the
stratification in the upper layers where convective fluctuations are in
the non-linear regime, and where radiation escapes the stellar surface.

The transition from diffusive to free-streaming radiative transfer ({\ie}, the
photosphere) is dealt with in great detail in conventional 1D stellar atmosphere
calculations, where nowadays $10^5$--$10^6$ wavelengths are used and non-LTE
effects are included when necessary. Convection, however, is described in the
mixing-length formulation which fails in the surface layers. The only
alternative that can result in realistic and self-consistent stratifications,
spectral line-shapes\cite{asplund:solar-Fe-shapes}, abundances
\cite{asplund:solar-Fe-abund}, {\etc}, are the convection simulations.
It has recently been shown that a simplified radiative transfer scheme for the
simulations, can reproduce the monochromatic solution, but for a small fraction
of the computational cost \cite{trampedach:SOS-Liv2002}. Apart from non-LTE
effects, the radiative transfer in the simulations can rival that of modern
1D atmosphere calculations, but with a realistic and consistent treatment
of convection.

The convection simulations are deeper than normal atmosphere models, since, in
general, convection turns adiabatic in deeper layers than where radiative
transfer turns diffusive. As a consequence, the convection simulations straddle
the regimes of a simple Saha {\eos} that includes molecules, and a
more sophisticated one that, however, can neglect molecules (other than H$_2$).
The desire to accurately match the convection simulations with interior models,
further promotes the concept of a stellar {\eos} that includes everything from
molecules at low temperature, interaction effects at high densities and 
relativistic effects at high temperatures. The present paper is an outline
of such work in progress.

\section{The MHD equation of state}

The MHD equation of state \cite{mhd2,mhd3} was conceived as part of the
International Opacity Project (OP) \cite{OP1,OP2}, of the late 1980's, to
calculate precise and accurate opacities for stellar envelopes. The emphasis
was on detailed quantum-mechanical calculations of the electron structure
of all ions of the astrophysically most important elements. The resulting
energy-levels and effective radii were used in the MHD equation of state
and the transition probabilities were used for calculation of absorption
coefficients for each state of each atom/ion. Combining the population of these
states, as found from the equation of state, with the absorption coefficients,
then results in the opacity; a self-consistent set of calculations.

The OPAL project \cite{rogers:OPAL-EOS,fr:OPAL-opac} is a parallel and similar
{\eos} and opacity effort, but
independent of MHD and OP, and pursued in the physical picture.

The key concept introduced in the MHD \eos, was that of occupation
probabilities, $w_i$, of a particular state, $i$
\cite{mhd1}. The Boltzmann-factors in
conventional calculations of the population of states, are based on the
assumption of an isolated particle and therefore ignore the presence
of neighboring particles. This results in the well-known 24\% un-ionized
hydrogen at the center of the Sun, in this simplistic picture. It is obvious
that the density at the Solar center leaves no room for hydrogen atoms.
Two methods were introduced to cut-off the summation of the otherwise divergent
partition function; one considered the available volume and the size of bound
states and assumed dense packing, the other used a lowering of the ionization
potential to spill bound states into the continuum. Both methods introduce
discontinuities into the \eos~and most implementations have lacked internal
consistency.

The MHD \eos, on the other hand, includes a realistic model of a physical
process that directly leads to ionization; ionization by repeated crossings
of Stark-manifolds by fluctuating electric fields, as measured by
\cite{pillet:field-ioni}. This leads to ionization when the amplitude of the
fluctuating field exceeds a threshold, $F_{\rm cr}$, for given state.
The probability of state $i$ being present in a plasma is therefore
\be{QFcr}
	w_i = Q(F_{{\rm cr},i}) = \int_0^{F_{{\rm cr},i}} P(F){\rm d}F\ ,
\ee
where $P(F)$ is the micro-field distribution. This provides us with a smooth,
self-consistent and physically plausible \eos.

A further advantage over previous work is the analytical derivatives, which
ensures smooth and consistent thermodynamic derivatives.

\section{Beyond the original MHD}

The improvements detailed below, are compared to the original MHD and to the
OPAL \eos \cite{rogers:OPAL-EOS}
in Figs.\ \ref{mhd_suncmp_Qdistr}--\ref{mhd_suncmp_Coul}. These plots show
pressure differences in the left-hand panel and $\gamma_1$ differences in the
right-hand panel. These quantities have been selected for their importance for
the hydrostatic structure and the sound-speed profile, respectively.
In these plots the solid line shows the difference (new MHD$-$original MHD) as
the solid lines, (OPAL$-$original MHD) as the gray lines. The dotted lines show
the original MHD (as published) with the $\tau$-function that was introduced to
curb the otherwise diverging (with density) Debye-H\"uckel term. This
$\tau$-factor was discussed by \cite*{rt:EOS-comp}, disputing its physical
foundation and showing that better agreement with OPAL is obtained using
$\tau=1$, as shown with the gray lines in
Figs.\ \ref{mhd_suncmp_Qdistr}--\ref{mhd_suncmp_Coul}. The solid, the dotted and
the gray lines are the same in all three plots. The dashed lines show the effect
of turning off some of the improvements, one at a time.

The OPAL {\eos} is an important comparison case, since Solar models based on it
show overall better agreement with helioseismic inversions
\cite{jcd-wd:osc-eos}, compared to MHD.
Looking more carefully at the outer layers of the Sun, using high degree
p-modes, models using the MHD {\eos} seems to be closer to the Sun in the
outer 7\%, than ones using OPAL \cite{dimauro:high-l-suncmp}. This region
includes the H-, He- and some
of the He$^+$-ionization zone, and covers the left half of each of the plots
in Figs.\ \ref{mhd_suncmp_Qdistr}--\ref{mhd_suncmp_Coul}. Deeper in the Sun
we therefore want better agreement with OPAL ({\ie}, the solid curve close
to the gray curve in Figs.\ \ref{mhd_suncmp_Qdistr}--\ref{mhd_suncmp_Coul}),
whereas the situation is less clear further out.

\subsection{Micro-field Distributions}
\begin{figure}
  \includegraphics[height=.5\textheight]{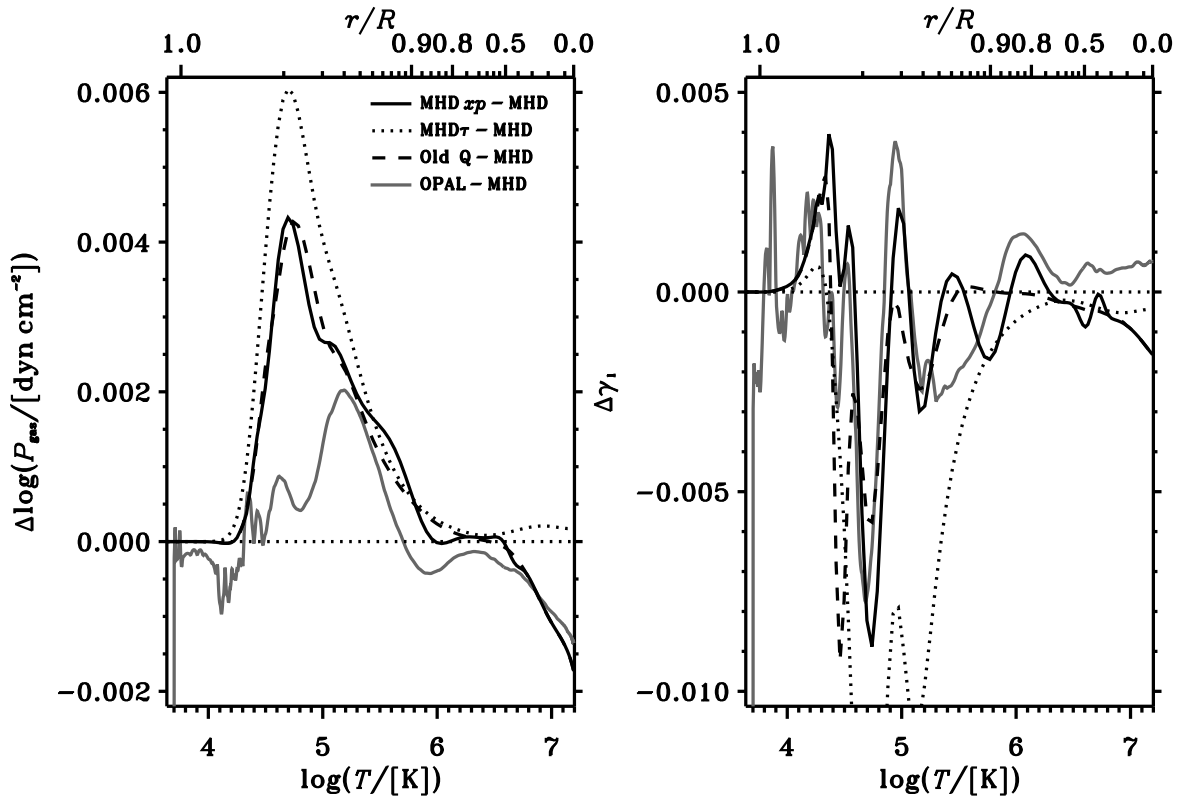}
  \caption{\label{mhd_suncmp_Qdistr}Differences between original MHD and:
           OPAL (gray), all changes (solid), all changes except old micro-field
		   distribution (dashed), along a Solar $\varrho/T$-stratification.}
\end{figure}

The original MHD formulation used linear fits to the Holtzmark distribution,
which assumes non-interacting particles. \cite*{Qmhd} introduced much improved
fits to a micro-field distribution model that accounts for particle interactions
through the Debye-H\"uckel theory. I have included this so-called $Q$-MHD
formulation and the effect under Solar circumstances can be seen in
Fig.\ \ref{mhd_suncmp_Qdistr}.
The effect is largest in the derivatives, where a lot of new structure is
introduced. This change will most likely have a larger effect on the opacity
since it changes the population of highly excited states that are often strong
absorbers.

\subsection{Quantum effects}

Apart from the effect on particle statistics, in the form of degeneracy,
there are at least two other effects of quantum mechanics that needs to be
included; quantum diffraction and exchange interactions. The former is a
consequence of Heisenberg's uncertainty relation and the realization that
particles are wave-packets of finite size. This results in finite
charge-densities and therefore avoids the short-range divergence of the
Coulomb potential. This effect replaces the $\tau$-factor mentioned above.

\begin{figure}
  \includegraphics[height=.5\textheight]{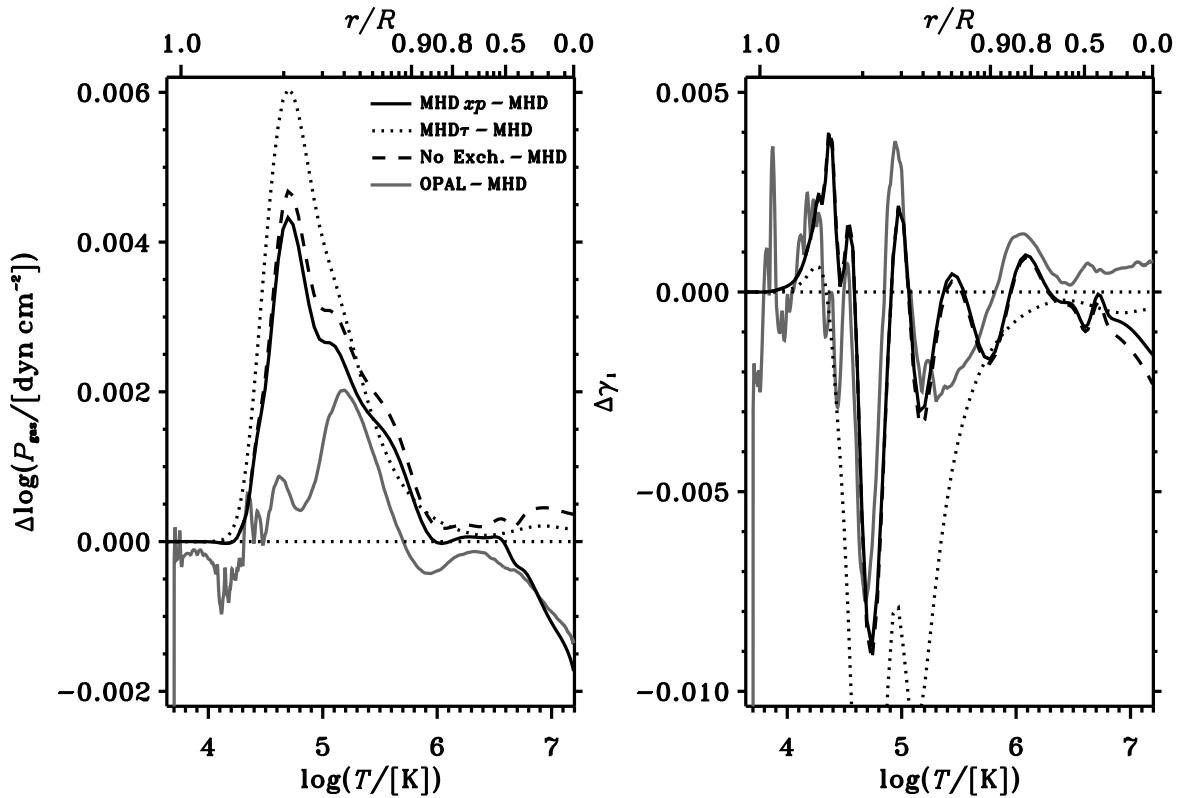}
  \caption{\label{mhd_suncmp_Exch}Differences between original MHD and:
           OPAL (gray), all changes (solid), all changes except for
		   exchange effects (dashed), along a Solar $\varrho/T$-stratification.}
\end{figure}

The exchange term arise from Pauli's exclusion principle between
identical particles; the wave-functions of two identical particles will
either overlap or repel each other, depending on their spin, thereby
changing their interaction energy compared to that of differing particles.
The effect of the first-order exchange-term \cite{kovetz:exchange}
is shown in Fig.\ \ref{mhd_suncmp_Exch}.

\subsection{Interactions with Neutral Particles}

In the original MHD {\eos} neutral particles were treated as hard spheres,
but only to first order in particle density, $N_i$. The second order term
was included in an approximate way through the $\Psi$-function discussed by
\cite*{rt:EOS-psi}. Apart from these approximations, the hard-sphere model
has its problems. First of all, it is undefined for high densities and can
disrupt the convergence of the free-energy minimization. Second, it is
un-physical and ignores the underlying forces.
A neutral atom has an extended electron distribution, and in
close encounters, another particle will dip into this electron wave-function
and begin to feel the charge of the nucleus which is no longer completely
screened. At high densities net-neutral particles will therefore have a
small effective charge. The nice thing about this model, is that it also
applies to partially stripped ions, and all particles are treated on an equal
footing. Under Solar circumstances this change has no discernible effect,
but it most likely will for cooler stars.

\subsection{Coulomb Interactions}
\begin{figure}
  \includegraphics[height=.5\textheight]{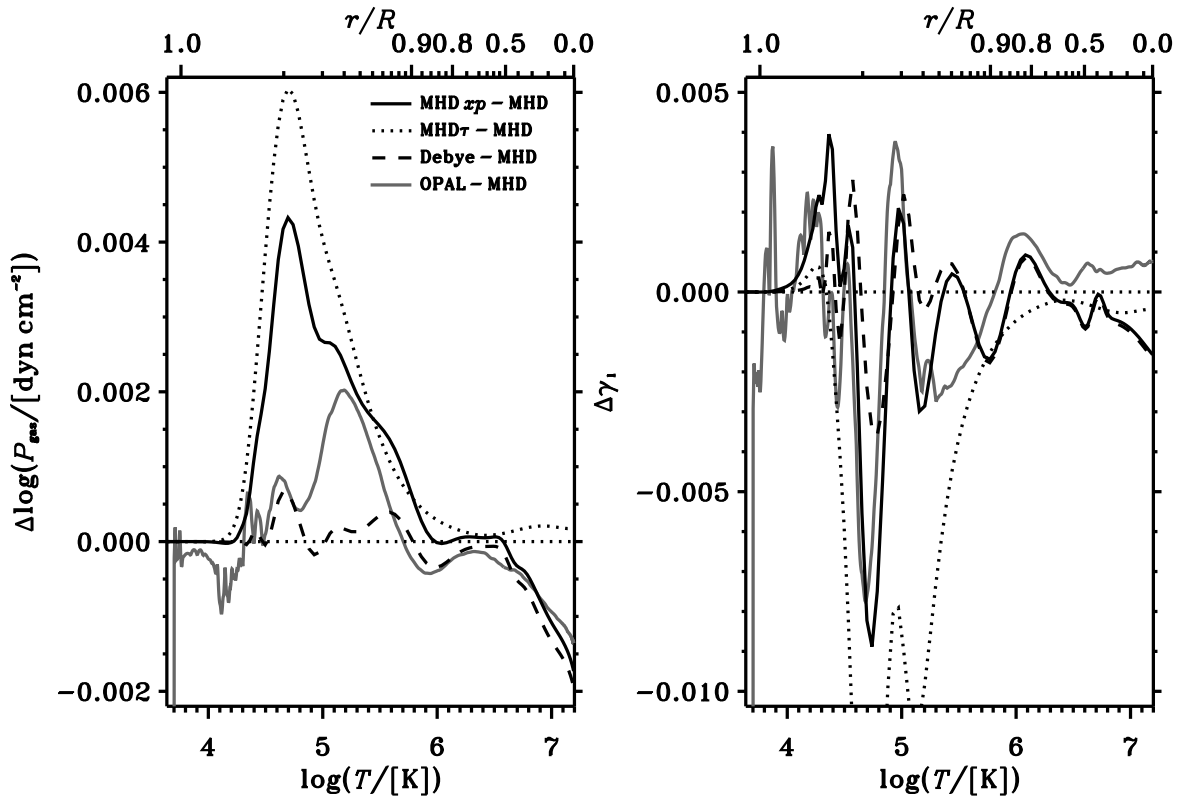}
  \caption{\label{mhd_suncmp_Coul}Differences between original MHD and:
           OPAL (gray), all changes (solid), all changes except pure
		   Debye-H\"uckel (dashed), along a Solar $\varrho/T$-stratification.}
\end{figure}

By far the largest change---at least in the Sun---arise from including
higher-order terms in the Coulomb interactions, beyond the Debye-H\"uckel
term, $F_{\rm DH}$. The original MHD had a $\tau$-correction to $F_{\rm DH}$
that turned out to be un-physical \cite{rt:EOS-comp},
and is now replaced by a correction-factor
based on Monte-Carlo simulations of the one-component-plasma
\cite{slattery:OCP-MC}, implicitly including many-body interactions. The effect
of this change can be seen in Fig.\ \ref{mhd_suncmp_Coul}; it is constrained to
the convection zone and peaks at a depth of about 7\,Mm, which is surprisingly
close to the surface.

\subsection{Additional Changes}

Relativistically degenerate electrons are included as detailed by
\cite*{GDZ:rel-e-MHD}. The effect is very small except in the adiabatic
exponent, $\gamma_1$, which is lowered appreciably in the inner half of the Sun.Relativistic effects are not included in the version of OPAL we have used
for comparison, but omitting it in the new MHD results in very good agreement
with OPAL for $\log T > 6.7$. The new version of OPAL includes relativistic
electrons, as well as a few other improvements \cite{rogers:newOPAL}.

Molecules are included by means of the fits to partition functions, compiled
by \cite*{irwin:mol-part-func}, augmented by a parametrized pressure
dissociation, based on the detailed treatment by MHD of the H$_2$- and
H$_2^+$-molecules. Molecules are selected from a list of 315 di-atomic and 99
poly-atomic molecules, depending on the elements included in the \eos~
calculation. This change will mostly affect the atmospheric opacities.

\section{Summary}

A new {\eos} is being developed, based on the MHD {\eos}, but with a number
of improvements that will hopefully bring it on par with the OPAL {\eos}.
Tables have not been computed yet, so calculations of Solar models with the
new {\eos} are not yet possible. Comparisons on a fixed $\varrho/T$-track
of a Solar model, as presented here, are encouraging, however.

%%%%%%%%%%%%%%%%%%%%%%%%%%%%%%%%%%%%%%%%%%%%%%%%
%% BACKMATTER
%%%%%%%%%%%%%%%%%%%%%%%%%%%%%%%%%%%%%%%%%%%%%%%%

\begin{theacknowledgments}
I am grateful to the organizers of the workshop, for the invitation to present
and for a fruitful two weeks. I would also like to thank Werner D\"appen for
full access to the original MHD-code, including rights to and help with
modifying it.
\end{theacknowledgments}

%% If the bibliography is produced without BibTeX comment out the

\bibliographystyle{aipproc}   % if natbib is available

%\bibliography{bibs/eos,bibs/opac,bibs/seism,bibs/conv,bibs/starmod}

\end{document}